\documentclass[aps,pre,twocolumn,floats,showpacs]{revtex4}
\usepackage{epsfig}
\usepackage{color}
\usepackage{graphicx}
\usepackage{caption}
\usepackage{amsmath, amssymb}
\usepackage{bbold}
\usepackage{bm}
\usepackage{latexsym}

\begin{document}
\newcommand{\hide}[1]{}
\newcommand{\tbox}[1]{\mbox{\tiny #1}}
\newcommand{\half}{\mbox{\small $\frac{1}{2}$}}
\newcommand{\sinc}{\mbox{sinc}}
\newcommand{\const}{\mbox{const}}
\newcommand{\trc}{\mbox{Tr}}
\newcommand{\intt}{\int\!\!\!\!\int }
\newcommand{\ointt}{\int\!\!\!\!\int\!\!\!\!\!\circ\ }
\newcommand{\eexp}{\mbox{e}^}
\newcommand{\bra}{\left\langle}
\newcommand{\ket}{\right\rangle}
\newcommand{\EPS} {\mbox{\LARGE $\epsilon$}}
\newcommand{\ar}{\mathsf r}
\newcommand{\im}{\mbox{Im}}
\newcommand{\re}{\mbox{Re}}
\newcommand{\bmsf}[1]{\bm{\mathsf{#1}}}
\newcommand{\beq}{\begin{equation}}
\newcommand{\eeq}{\end{equation}}
\newcommand{\bea}{\begin{eqnarray}}
\newcommand{\eea}{\end{eqnarray}}
\definecolor{red}{rgb}{1,0.0,0.0}

\title{Complexity of two-level systems}

\author{Imre Varga}
\affiliation{Department of Theoretical Physics, Institute of Physics, Budapest University of Technology and
Economics, Műegyetem rkp. 3., H-1111 Budapest, Hungary}
\date{\today}
\begin{abstract}
Complexity of two-level systems, e.g. spins, qubits, magnetic moments etc, are analysed based on the so-called correlational entropy in the case of pure quantum
systems and the thermal entropy in case of thermal equilibrium that are suitable quantities essentially free from basis dependence. The complexity is defined as the
difference between the Shannon-entropy and the second order R\'enyi-entropy, where the latter is connected to the traditional participation measure or purity.  It is shown that the system 
attains maximal complexity for special choice of control parameters, i.e. strength of disorder either in the presence of noise of the energy states or the presence of disorder in the 
off diagonal coupling. It is shown that such a noise or disorder dependence provides a basis free analysis and gives meaningful insights. We also look at similar entropic complexity 
of spins in thermal equilibrium for a paramagnet at finite temperature, $T$ and magnetic field $B$, as well as the case of an Ising model in the mean-field approximation. As a result
all examples provide important evidence that the investigation of the entropic complexity parameters help to get deeper understanding in the behavior of these systems.
\end{abstract}

\maketitle

\section{Introduction}
There has been much interest over the past decade in introducing quantities in order to characterise the complexity of a given system~\cite{comp_book}. One of the pioneering works 
is the one by L\'opez-Ruiz, Mancini and Colbet~\cite{LMC} often referred to as the LMC complexity. This ideas has been extended and applied in a number of later 
publications~\cite{Lopez}. In parallel a different line of research has been developed using appropriate parameters~\cite{VargaPipek} for the analysis of probability distributions 
and it has been found to be closely related to the LMC complexity~\cite{Lopez} but the definition of the so-called {\it structural entropy} is mathematically well defined.
The term structural was introduced in order to separate that from the extension entropy since the sum of these two non--negative quantities always sum up to the well-known
Shannon-entropy. We also have to mention a more recent suggestion, where the so-called information capacity~\cite{Landauro} has been calculated for a number of systems. 
Apparently its major role is to locate a point in the parameter space where complexity is maximal showing the intermediate state between order and chaos, or order and disorder. 
As it will be clear in the present paper, as well, the quantity we propose shows similar advantages and it is mathematically more founded as compared to the previous ones. 

The analysis of especially quantum systems suffers from the problem that eigenstates are basis dependent. Sometime that ambiguity is obvious whenever there exist a preferential
basis. On the other hand there has been some previous work trying to consider basis independent description of quantum states. Along this line some decades ago the quantity 
termed as the correlational entropy has been introduced~\cite{Sokolov88}. The idea was to build on a basis independent definition of (generalized) entropies. Sokolov and coworkers 
in several papers have shown the possible applicability of the so-called correlational entropy. Here we show, that based on these grounds one can in principle elaborate further to investigate 
for instance the complexity of two-level systems (TLS).

\section{The concept of correlational entropy and the derivation of the complexity therefrom}
In case of a system described by a hamiltonian, $H(\lambda )$ depending on a parameter $\lambda $ that may be random with an appropriately chosen probability density function (PDF). 
The solution of the eigenvalue equation
\beq
H(\lambda)|\alpha;\lambda\rangle = E_{\alpha}(\lambda)|\alpha;\lambda\rangle,
\eeq
where the eigenvectors can be expressed as a linear combination of the original, orthonormal, computational basis $|k\rangle$
\beq
|\alpha;\lambda\rangle = \sum_k c_k^{\alpha}(\lambda)|k\rangle.
\eeq
The elements of the density matrix $\rho^{(\alpha)}$ of a pure eigenstate $|\alpha;\lambda\rangle$,  can be expressed using these coefficients as
\beq
\rho_{k,k'}^{(\alpha)}(\lambda)=\langle k|\alpha;\lambda\rangle\langle\alpha;\lambda| k' \rangle =c_k^{\alpha}(\lambda)c_{k'}^{\alpha}(\lambda)^*.
\eeq
The density matrix is hermitian and for the pure eigenstate we have 
\beq
\trc \rho^{(\alpha)}=\trc [\rho^{(\alpha)}]^2 = 1,
\label{eq.tr}
\eeq
However, following~\cite{Sokolov88} we assume that $\lambda$ is a random variable with a given PDF. Hence the average of the density matrix will be calculated 
over the PDF, $P(\lambda)$ of this variable $\lambda$
\beq
\rho_{k,k'}^{(\alpha)}=\overline{\rho_{k,k'}^{(\alpha)}(\lambda)} \equiv \int d\lambda P(\lambda)  c_k^{\alpha}(\lambda)c_{k'}^{\alpha}(\lambda)^*,
\eeq
where the density matrix remains hermitian but no longer keep the pure state property, Eq.~(\ref{eq.tr}). 

The eigenstates of the density matrix can be characterised by entropies, i.e. the von Neumann or Shannon entropy
\beq
S(\alpha)=-\trc \left \{\rho^{(\alpha)}\ln \rho^{(\alpha)}\right \}
\label{entropy}
\eeq
and an apropriate generalization, the special R\'enyi entropy~\cite{Renyi} of order 2 that is directly connected to purity and the so-called IPR, the inverse participation ratio
\beq
R_2(\alpha)=-\ln \trc \left \{\left[\rho^{(\alpha)}\right ]^2\right\}.
\label{purity}
\eeq
Both of these entropies can be called correlational in the sense introduced by Sokolov {\it et al.}~\cite{Sokolov88} that has been successfully applied in cases of
many body chaos mainly in nuclear physics. Hereby we wish to import this idea into quantum two--level systems (TLS) that are the essential models
of quantum computing via the notion of the qubit. We wish to investigate its applicability if the randomness arises from the inevitable noise induced by the environment. 
Furthermore, we wish to introduce a parameter, statistical entropic complexity measure~\cite{new1} which has been successful for many other purposes~\cite{VargaPipek}.
We have to mention that R\'enyi entropies have already been applied as possible generalizations of measures of complexity~\cite{Renyi2}.

The parameter that we wish to calculate which has been used for many previous examples is the so-called structural entropy but from now on it will be
termed as entropic complexity, i.e. $S_C$ defined using definitions Eqs.~(\ref{entropy}, \ref{purity}) as
\beq
S_C(\alpha)=S(\alpha)-R_2(\alpha)
\label{sstr}
\eeq
Analogous versions have been successfully applied in various cases~\cite{VargaPipek}. It is also very similar to the so-called 
LMC complexity parameter but is well-founded and has roots back to localization properties and hence $S_C$ has been useful as $S_{str}$ rather describing the shape of various PDF-s. 
It is a non-negative quantity and the more the PDF deviates from a uniform distribution the larger it becomes, hence its usage describing the shape of a PDF.
Indeed the LMC parameter and our $S_{str}$ have been shown to be practically equivalent~\cite{Lopez, new1}.

\section{The model of two level systems (TLS)}
The simplest model of a qubit is a two--level system defined by the following Hamiltonian
\beq
H=\frac{1}{2}(\varepsilon -\lambda)\sigma_z + V\sigma_x,
\label{TLS-Ham}
\eeq
where the $V$ describes the strength of mixing of the levels defined as the diagonal energies. The $\sigma_x$ and the $\sigma_z$ are the usual Pauli-matrices.
The energy parameters, $\varepsilon$, $\lambda$ and $V$ define several, different possible variants. Letting $\varepsilon=0$ provides two levels, $\pm\lambda$ symmetrical
about $E=0$ with a mixing rate $V$. Usually $V=1$ can be taken as the unit of energy if not specified otherwise. 

In general the Hamiltonian, Eq.~(\ref{TLS-Ham}) can be diagonalised for a given value of $\lambda$ using a $2\times 2$ unitary rotation with an angle $\phi$ defined as
\beq
\sin\phi = \frac{2V}{\Delta(\lambda)}, \qquad \cos\phi = \frac{\varepsilon-\lambda}{\Delta(\lambda)},
\label{cands}
\eeq
where
\beq
\Delta (\lambda) = \sqrt{(\varepsilon-\lambda)^2+4V^2}
\eeq
is the level spacing. The density matrix for the two eigenstates is
\beq
\rho^{(\pm)}(\lambda)=\frac{1}{2}\left [\mathbb{1}\pm (\sigma_x\sin\phi+\sigma_z\cos\phi) \right],
\eeq
where $\mathbb{1}$ represents a $2\times 2$ identity matrix.
Upon averaging over the random variable $\lambda$ or $V$, we arrive to 
\beq
\rho^{(\pm)} = \frac{1}{2}[\mathbb{1}\pm (\sigma_x s+\sigma_z c) ],
\eeq
where $s=\overline{\sin\phi}$ and $c=\overline{\cos\phi}$, are averaged quantities which do not satisfy $c^2+s^2=1$. The eigenvalues of this averaged density matrix are
\beq
\rho_\nu=\frac{1}{2}(1+\nu r), \qquad r=\sqrt{s^2+c^2}, \qquad \nu=\pm 1.
\eeq
Therefore the entropy is the same for both eigenstates of the density matrix. Using the entropic definition of the complexity parameter, Eq.~(\ref{sstr}) we obtain
\beq
S_{C}=-\frac{1+r}{2}\ln\frac{1+r}{2}-\frac{1-r}{2}\ln\frac{1-r}{2}+\ln\frac{1+r^2}{2}.
\label{sstr_TLS}
\eeq
Hereby we also normalize to $\ln 2$ as for the two-state problem the maximum of the entropy is $\ln 2$.

We have to point out that as far as the complexity parameter is concerned, identical results can be obtained even if the TLS is considered as a half spin in a random field given as
\beq
H=\vec\sigma \cdot {\bf n},
\eeq
where $\bf n$ is a random unit vector. The density matrix in this case is written as
\beq
\rho^{(\pm)}=\frac{1}{2}(\mathbb{1}\pm \vec\sigma \cdot \bf n).
\eeq

\subsection{Landau-Zener system}
First we calculate the complexity parameter for the ideal, non-random case with $\varepsilon=0$. This is a model known as the Landau-Zener-St\"uckelberg-Majorana model~\cite{LZ}
(LZSM) which will be the starting basis for our analysis, as well. 

\subsubsection{Varying levels}
In this case the TLS Hamiltonian can be written as
\beq
H(t)=\alpha t\sigma_z+V\sigma_x=
\left( {\begin{array}{cc}
\alpha t &      V       \\
     V     & -\alpha t  \\
  \end{array} } \right),
\label{LZ-diag}
\eeq
where the system is governed by a fictitious time parameter $T_0=\varepsilon/\alpha$, such that $\alpha t=\varepsilon t/T_0$, therefore any quantity can be
calculated as a function of $x=\gamma t/T_0$ with $\gamma =\varepsilon/V$. The eigenvalues are given as $E_{\pm}(x) =\pm V\sqrt{x^2+1}$, 
and we choose $V=1$ as the unit of the energy. Obviously for $x=0$ the gap is the smallest, but it is still nonzero, $E_{+}-E_{-}=2V$. The eigenvectors can be given as 
$|+\rangle = c_+|1\rangle + |0\rangle$ and $|-\rangle = |1\rangle + c_-|0\rangle$ in the computational basis with the additional constraints that
$\langle \pm|\pm\rangle=1$ and $\langle \mp |\pm\rangle=0$. Hence for the eigenstates one obtains (before normalization)
\beq
c_+^{(di)}(x)= \sqrt{1+x^2}+x =-c_-^{(di)}(x).
\label{c_diag}
\eeq
Therefore our complexity parameter turns out to be the same for both eigenstates, $|\pm\rangle$ which reads as
\beq
S_{C}=-\frac{c_{\pm}^2}{1+c_{\pm}^2}\ln c_{\pm}^2 + \ln\frac{1+c_{\pm}^4}{1+c_{\pm}^2}.
\label{sstr_LZ}
\eeq
\begin{figure}
\epsfxsize=8.4cm
\leavevmode
\epsffile{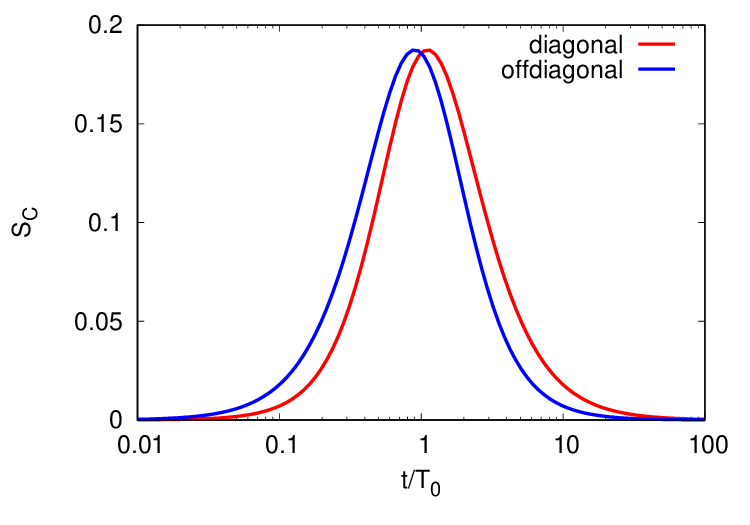}
\caption{Complexity of any of the two states (ground or excited) of the Landau-Zener problem for varying either the diagonal or the off-diagonal energy parameters.}
\label{fig:LZ}
\end{figure}

\subsubsection{Varying coupling}

Now, let us turn to a similar TLS problem, but with fixed diagonal parameters and linearly changing off-diagonal ones.
\beq
H(t)=\varepsilon \sigma_z+\alpha t\sigma_x=
\left( {\begin{array}{cc}
\varepsilon &    \alpha t     \\
    \alpha t   & -\varepsilon \\ 
  \end{array} } \right),
\label{LZ-offd}
\eeq
where the system is governed by a fictitious time scale $T_0=V/\alpha$, such that $\alpha t=V t/T_0$, therefore any quantity can be
calculated as a function of $x=\delta t/T_0$ with $\delta =V/\varepsilon$. The eigenvalues are given as $E_{\pm}(x)=\pm\varepsilon\sqrt{1+x^2}$, 
where without loosing the generality, we will assume $\varepsilon=1$ as the unit of energy. Using the same procedure as in the previous
subsection we obtain the following eigenstate components (before normalization)
\beq
c_+^{(od)}(x)=\frac{\sqrt{1+x^2}+1}{x}=-c_-^{(od)}(x).
\label{c_offd}
\eeq
The formula for the complexity parameter is the same for both states and can be given as Eq.~(\ref{sstr_LZ}) as for the diagonal case. 
Both diagonal and off-diagonal $x$, i.e. $t$ dependences are shown in Fig.~(\ref{fig:LZ}). 

From the figure we can see that the maximal complexity is reached very close to $x=1$, i.e. when $t=\gamma T_0$ and $t=\delta T_0$ for the two cases, basically using $\gamma=\delta=1$
without any loss of generality. However, for the case when the diagonal parameter changes with $t$ this point is 10\% higher ($x_c^{(di)}=1.110668\dots$) and when the off-diagonal parameter 
changes with $t$, then the position of the maximum is  10\% lower ($x_c^{(od)}=0.900359\dots$) than unity. There is also a remarkable symmetry on a logarithmic scale of $x$ about the maximum 
for both curves, which is a direct consequence of the symmetry between Eq.~(\ref{c_diag}) and Eq.~(\ref{c_offd}) about $x=1$. Such a behavior is new that shows a markedly outstanding complexity 
at roughly $t=T_0$, while the $t=0$ or rather $t\ll T_0$ and $t\to\infty$ or $t\gg T_0$ are  trivial and apparently interchangeable as far as the complexity is concerned. Due to the symmetry between
Eq.~(\ref{c_diag}) and Eq.~(\ref{c_offd}) about $x=1$ on a logarithmic scale, i.e. $c_+^{(di)}(x)=c_+^{(od)}(1/x)$, Eq.~(\ref{c_diag}) at $x^{(di)}$ and Eq.~(\ref{c_offd}) at $x^{(od)}$ we get exactly the 
same coefficients (after normalization), $c_+=0.93358\dots$. In other words $|+\rangle$ with maximal complexity has a 87.16\% population on $|1\rangle$ and 12.84\% population on $|0\rangle$. 
Likewise the state $|-\rangle$ has 12.84\% population on $|1\rangle$ and 87.16\% population on $|0\rangle$. Using a Bloch-sphere representation
\beq
|\pm\rangle = \cos\frac{\theta_{\pm}}{2}|0\rangle + e^{i\varphi_{\pm}} \sin\frac{\theta_{\pm}}{2}|1\rangle
\eeq
for the states with maximal complexity we have $\varphi_+=0$ and $\theta_+ =2.40856\dots = 138^{\circ}$ and $\varphi_-=\pi$ and $\theta_- =0.73302\dots = 42^{\circ}$. 
The $\theta$ angles measured in degrees are remarkably sharp values. 

\subsection{TLS with fluctuating parameters}
As for the original TLS problem we will calculate the complexity parameter for several cases using binary and uniform disorder both for the fluctuation of the levels $\lambda$
or the fluctuation of the coupling parameter $V$. Note that parameter $\varepsilon$ bares an important role, as for diagonal disorder, a random distribution of $\lambda$ produces
different behavior depending on the fluctuations of $\lambda$ compared to $\varepsilon$. As long as $\varepsilon=0$, our starting model is the original LZSM model and will see either
the same behavior as that obtained using a deterministic variation of the diagonal parameter as a function of the fictitious time, $t$ or we may observe departures therefrom. Hence the
major effect is expected as $W\approx \varepsilon$ and/or $W\approx 2V$, where $W$ describes the variance of the distribution of $P(\lambda)$ and $2V$ is the  smallest level separation of
the original LZSM model in case of $\varepsilon=0$, i.e. describing the minimal mixing strengths.

\subsubsection{Binary distribution}

First we will take the case of binary distributions. In case if the diagonal parameter $\lambda$ in Eq.~(\ref{TLS-Ham}) is chosen from a PDF described by
$P(\lambda)=\half [\delta(\lambda-W)+\delta(\lambda+W)]$. As long as we consider the TLS Eq.~(\ref{TLS-Ham}) using $\varepsilon=0$, we may expect the random case to produce identical 
complexity as for the ideal LZSM system. Calculating the values of $s=\overline{\sin\phi}$ and $c=\overline{\cos\phi}$ for this special case and obtain
\bea
c&=&\frac{1}{2}\left [\frac{\chi -1}{\sqrt{(\chi -1)^2+\tau^2}}+\frac{\chi +1}{\sqrt{(\chi +1)^2+\tau^2}} \right ], \\
s&=&\frac{\tau}{2}\left [\frac{1}{\sqrt{(\chi -1)^2+\tau^2}}+\frac{1}{\sqrt{(\chi +1)^2+\tau^2}} \right ],
\eea
where $\chi=\varepsilon/W$ and $\tau=2V/W$ describe the ratio of the original parameters $\varepsilon$ and $V$ compared to the measure of disorder $W$. Especially for the case
of $\varepsilon=0$, when $\chi=0$, these values simplify to $c=0$ and $s=\tau /\sqrt{1+\tau^2}$, and using them substituting in Eq.~(\ref{sstr_TLS}) we obtain exactly the same curves as that 
depicted in Fig.~\ref{fig:LZ}, which is also shown in Fig.~\ref{fig:combin}. 

\begin{figure}
\epsfxsize=8.4cm
\leavevmode
\epsfig{file=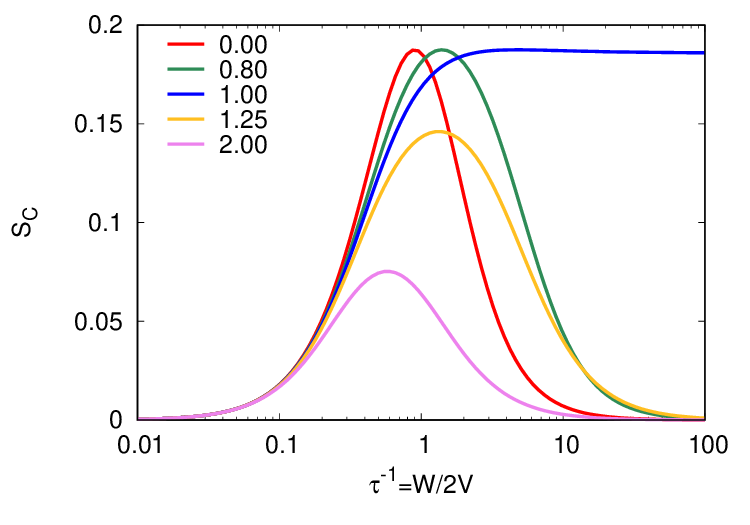,width=8.2cm}
\caption{Complexity of levels randomly distributed according to a binary--distributioon. The continuous curves are parametrized according to $\chi$.}
\label{fig:bin-diag}
\end{figure}

\begin{figure}
\epsfxsize=8.4cm
\leavevmode
\epsfig{file=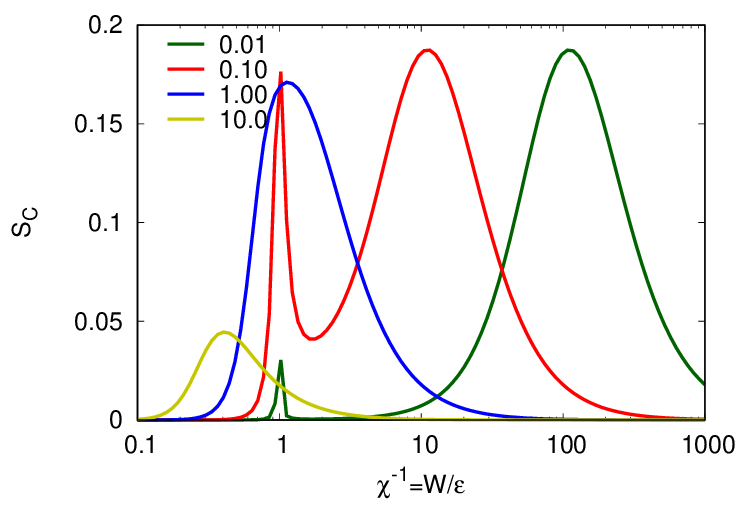,width=8.2cm}
\caption{Complexity of levels randomly distributed according to a binary--distribution  as a function of $\chi^{-1}=W/\varepsilon$. The continuous curves are parametrized according 
to $\zeta=\tau/\chi=2V/\varepsilon$. Note the second maximum of the curves for $\zeta\ll 1$ are located at values of $\chi^{-1}_c$ with $\chi^{-1}_c\zeta\approx 1$.}
\label{fig:bin-diag2}
\end{figure}

However, for a fixed value of $\varepsilon\neq 0$, we find a family of curves parametrized by $\chi=\varepsilon/W$ as a measure of disorder as depicted in Fig.~\ref{fig:bin-diag}.
Next we will present the complexity of the system as a function of parameter $\chi=\varepsilon/W$ and parametrized by the parameter $\tau=2V/W$. This is shown in Fig.~\ref{fig:bin-diag2}.

Considering randomly distributed level coupling in the case of a TLS using the following PDF, $P(V)=\half\left [\delta(V-V_0)+\delta(V+V_0) \right ]$ we obtain 
$s=0$ and $c=1/\sqrt{1+\kappa^2}$, with $\kappa=2V_0/\epsilon$ after setting the diagonal elements in Eq.~(\ref{TLS-Ham}) as $\epsilon$. In this case $r=c$ 
and so the function Eq.~(\ref{sstr_TLS}) gives identical result as the one depicted in Fig.~\ref{fig:LZ} for the case of the LZSM problem with varying off-diagonal 
matrix elements (see Eq.~(\ref{LZ-offd})). The latter comparison is presented in Fig.~\ref{fig:combin}.

\subsubsection{Box distribution}

\begin{figure}
\epsfxsize=8.4cm
\leavevmode
\epsfig{file=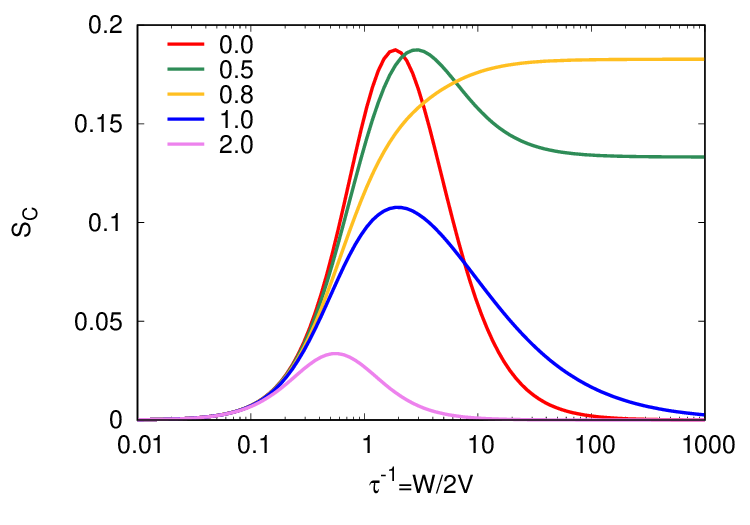,width=8.2cm}
\caption{Complexity of levels randomly distributed according to a box--distributioon. The continuous curves are parametrized according to $\chi$.}
\label{fig:box-diag}
\end{figure}

\begin{figure}
\epsfxsize=8.4cm
\leavevmode
\epsfig{file=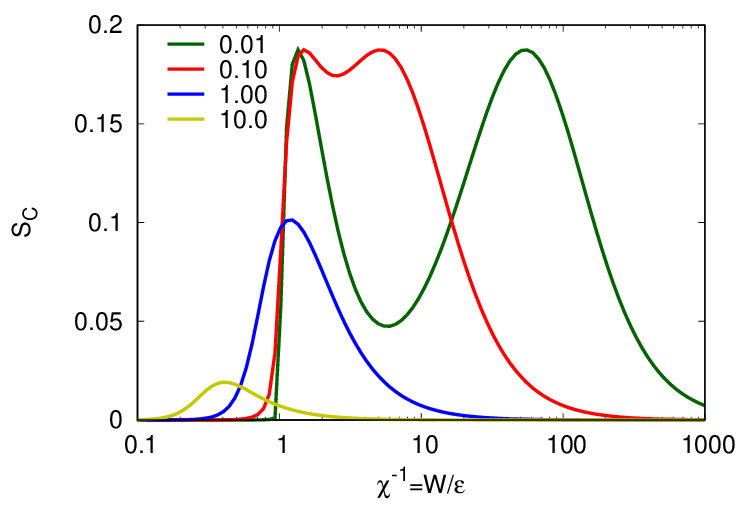,width=8.2cm}
\caption{Complexity of levels randomly distributed according to a box--distribution as a function of $\chi^{-1}=W/\varepsilon$. The continuous curves are parametrized according 
to $\zeta=\tau/\chi=2V/\varepsilon$. Note the second maximum of the curves for $\zeta\ll 1$ are located at values of $\chi^{-1}_c$ with $\chi^{-1}_c\zeta\approx 5$.}
\label{fig:box-diag2}
\end{figure}

\begin{figure}
\epsfxsize=8.4cm
\leavevmode
\epsfig{file=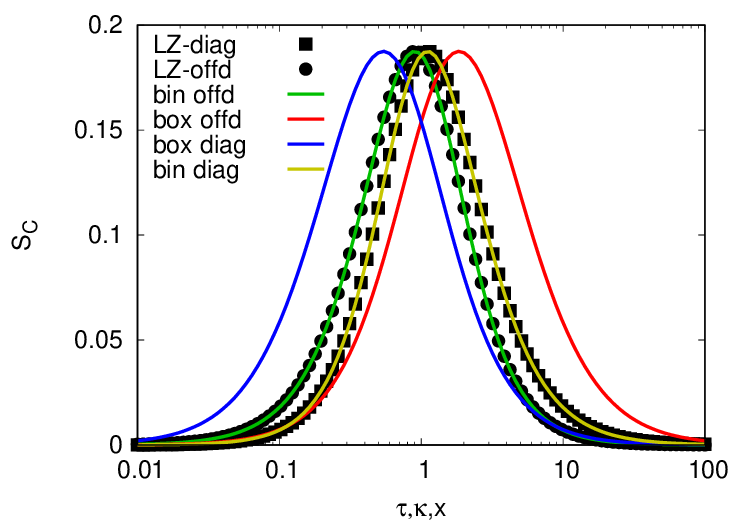,width=8.2cm}
\caption{Complexity of TLS with deterministic Landau-Zener problem and some special random binary and box distributions of the diagonal (using $\varepsilon=0$ in (\ref{TLS-Ham})) 
and off-diagonal (using $\lambda =0$ in (\ref{TLS-Ham})) parameters. The symbols show the behavior of the case of the LZSM problem already depicted in Fig.~\ref{fig:LZ}. The maximum
of the complexity for the box distribution of the parameter $V$ is reached at $\kappa_c^{box}\approx 1.85$, and the maximum for binary distributed $V$ is reached 
at $\tau_c^{bin}\approx 0.90$. The curve with random diagonal values of $\lambda$ drawn from a box distribution has a maximum at $\tau_c^{box}\approx 0.54$
and the maximum of the complexity for binary distributed diagonal elements $\lambda$ is reached at $\kappa_c^{bin}\approx 1.11$. 
Note that $\tau_c^{bin}\kappa_c^{bin}\approx \tau_c^{box}\kappa_c^{box}\approx 1$.}
\label{fig:combin}
\end{figure}

Now we consider noisy parameters chosen from a box distribution with finite width, $2W$ and zero mean. First we introduce noise in the levels of the TLS using
the PDF of the form $P(\lambda)=1/2W$ whenever $-W\leq\lambda\leq W$. From the parameters, $\varepsilon$, $W$, and $V$ we may introduce $\tau=2V/W$
and $\chi=\varepsilon/W$ just like in the case of binary noise. Calculating the values of $s=\overline{\sin\phi}$ and $c=\overline{\cos\phi}$, we get
\bea
c&=&\frac{1}{2}\left [\sqrt{(\chi+1)^2+\tau^2} - \sqrt{(\chi-1)^2+\tau^2} \right ], \\
s&=&\frac{\tau}{2}\ln\left [\frac{\sqrt{(\chi-1)^2+\tau^2} +\chi -1}{\sqrt{(\chi+1)^2 +\tau^2} +\chi +1} \right ]
\eea
Like in the previous case, here we can plot a family of $S_C$ curves as a function of $\tau$ parametrized by $\chi$. This is given in Fig.~\ref{fig:box-diag}.
As before we will present the complexity of the system as a function of parameter $\chi=\varepsilon/W$ and parametrized by the parameter $\tau=2V/W$. This is shown in Fig.~\ref{fig:box-diag2}.
There is also a special case of $\varepsilon=0$ when $\chi=0$ that yields $c=\overline{\sin\phi}=0$ and $c=\overline{\cos\phi}=\tau/2\ln [(\tau^2+2-2\sqrt{\tau^2+1})/\tau^2]$. The resulting
curves are depicted in Fig.~\ref{fig:combin}.

Finally let us turn to the case of random variation of the level coupling $V$ according to a box distribution, i.e $P(V)=1/2V_0$ if $|V|\leq V_0$, hence $2V_0$ is the width of the
distribution of $P(V)$. In this case we have only a single possibility, as the value of $s=\overline{\sin\phi}$ vanishes and $c=\overline{\cos\phi}=\ln(\sqrt{1+\kappa^2}+\kappa)/\kappa$, 
with $\kappa=2V_0/\epsilon$. The resulting complexity curve is shown in Fig.~\ref{fig:combin} together with the case when the
level couplings are drawn from a binary PDF. In addition the original Landau-Zener curve for off-diagonally varying coupling is also shown to be perfectly
identical to the one with binary distribution. As for the random variation of parameter $V$ using a box distribution, the system reaches a maximal
complexity for the value of $\kappa_c=1.848578\dots$.
As one can see the combination of the Landau-Zener curves and the simplest binary and box distributed the diagonal and the off--diagonal elements produce
interesting results. These are summarized in Fig.~\ref{fig:combin}. There is a remarkable correspondence and symmetry between these curves especially when 
plotted in the log of the parameter describing the complexity of the systems. The maxima seem to be symmetrical about unity which means that that their positions are 
the inverse of the ones on both sides. The analytical properties of the deterministic LZSM problem described above and shown in Fig.~\ref{fig:LZ} are presented 
as dots in Fig.~\ref{fig:combin}.

\section{A system of spins at finite temperature}

Hereby we wish to extend our complexity analysis of a system of two-level systems or spins at nonzero temperature and possibly at nonzero magnetic field.
Under the effect of both the temperature and magnetic field we may expect the following scenario. As long as the temperature is small enough and the magnetic
field is strong enough the spins more-or-less are expected to be aligned forming a basically ordered structure, whereas for large enough temperature and weak
magnetic field the spins are almost randomly oriented, hence there exist an intermediate regime of these parameters where the spins are neither ordered nor perfectly 
random and rather show a more complex nature therefore any complexity measure should show a maximum, while in the two extremes the complexity measure is expected to vanish. 

\subsubsection{General paramagnetic}
\begin{figure}
\epsfxsize=8.4cm
\leavevmode
\epsfig{file=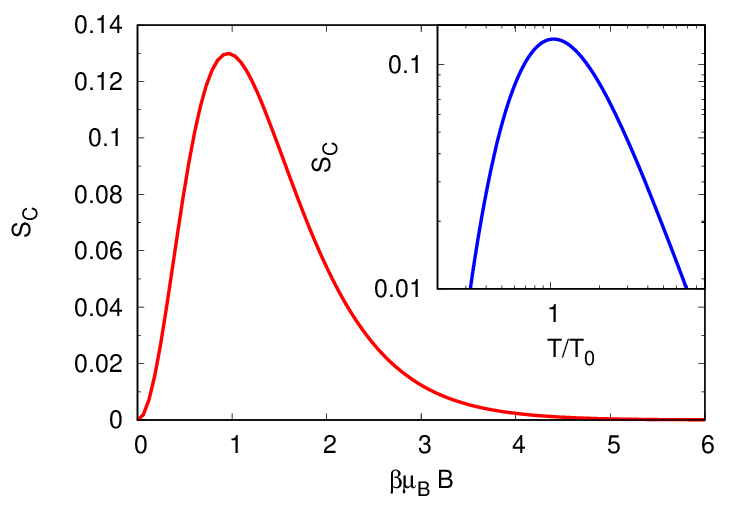,width=8.2cm}
\caption{Complexity of half-spins in the  paramagnetic phase. Note that the complexity is maximal roughly for $\mu_B B/k_B T_0=1$}
\label{fig:paramag}
\end{figure}

In this section we investigate the thermal distribution of a general system of spins and its complexity as a function of temperature $T$ and magnetic field $B$. We would like to study 
a well-known physical scenario of random magnetic moments represented by half-spins~\cite{Gould}. The energy of the spins in magnetic field, $B$ can be $\varepsilon=\pm\mu_B B$ 
($\mu_B$ is the Bohr magneton). Hence in thermodynamic equilibrium the magnetic moments at temperature $T$ and in the magnetic field $B$ will have a partition function per spin 
(from now on every quantity will be normalized to one spin):
\beq
Z = e^{\beta\varepsilon}+e^{-\beta\varepsilon}=2\cosh (\beta\varepsilon)
\eeq
with $\beta=\frac{1}{k_B T}$, therefore the occupation of any spin is
\beq
p=\frac{1}{Z}\exp(\beta\varepsilon)=\frac{e^{\beta\varepsilon}}{2\cosh (\beta\varepsilon)}
\eeq
hence the complexity of the system of half spins in a magnetic field $B$ at temperature $T$ in the paramagnetic phase is given as
\beq
S_C=-[p\ln (p)+(1-p)\ln (1-p)]+\ln (p^2+(1-p)^2).
\label{sstrB}
\eeq
The form of this function as a function of both $T$ and $B$ are presented in Fig.~\ref{fig:paramag}. It is clear that at $T_0=\mu_B B/k_B$, 
$B_0=k_B T_0/\mu_B$ the complexity parameter is maximal but at the extremes of vanishing $T$ and $B$ or infinite $T$ and $B$ this parameter should vanish, as well.
As Fig.~\ref{fig:paramag} shows the complexity clearly vanishes as the parameter $x=\mu_B B/k_B T$ either vanishes, $x\to 0$ or goes to infinity, $x\to\infty$. 
However, close to $x=1$ we see a maximum of complexity meaning the presence of competing effects of thermal fluctuations and ordering due to the external field.

\subsubsection{Ising model at mean-field level}

Next we take the example of a physical system of interacting spins. We look at the classical Ising problem in higher dimensions and here we will investigate the low temperature,
ferromagnetic phase
\beq
H=-J\sum_{\langle i,j\rangle}\sigma_i\sigma_j - \mu B\sum_i\sigma_i,
\eeq
where $\sigma_i=\pm 1$. Here $J$ is the interaction between nearest neighbor spins and $B$ is the external magnetic field. The coupling to the external field
is via $\mu=g\mu_B/2$, $\mu_B$ being the Bohr-magneton and $g$ is the giromagnetic ratio. In order to gain overall insights of the 
behavior of a system of spins under finite $B$ and $T$ we turn to its mean-field approximation~\cite{Arovas}, where
\beq
\sigma_i\sigma_j\approx \langle\sigma_i\rangle\langle\sigma_j\rangle + (\sigma_i-\langle\sigma_i\rangle)\sigma_j + (\sigma_j-\langle\sigma_j\rangle)\sigma_i 
\eeq
dropping negligible fluctuations, i.e. $(\sigma_i-\langle\sigma_i\rangle)(\sigma_j-\langle\sigma_j\rangle)\approx 0$. The magnetization per spin is
$m=\langle\sigma_i\rangle=\langle\sigma_j\rangle$. Then the minimization of the free energy yields the following equation to be solved for the magnetization 
$m$~\cite{Arovas}
\beq
m=\tanh \left[\beta (\varepsilon_B+mJz)\right ],
\label{eq.msc}
\eeq
where $z$ is the coordination number, $\beta=1/k_B T$ and $\varepsilon_B=\mu_B B$. This equation is called the Curie-Weiss equation.
Within this model we see a ferromagnetic ordering due to the coupling between adjacent spins which is disturbed by thermal fluctuations, therefore as $T\to 0$ 
ferromagnetic order dominates and above a certain temperature, $T_c$ disorder due to thermal fluctuations overcomes and the system loses its ferromagnetic
state. This is true especially in the absence of magnetic field, $B=0$, where we know that for $\beta Jz>1$ there are three solutions for Eq.~(\ref{eq.msc}), $m=0$, 
$m=\pm m_0$. Right at $\beta_c Jz=1$ we have the critical point, hence 
\beq 
k_BT_c=zJ,  \qquad \beta_c=1/(zJ).
\eeq
Therefore when $m=0$ the system is in the paramagnetic phase and for $m=\pm m_0$ it is in a ferromagnetic order and the critical point is located at $T_c$ named as the 
so--called Curie-temperature above which a ferromagnetic system turns paramagnetic. 

\begin{figure}
\epsfxsize=8.4cm
\leavevmode
\epsffile{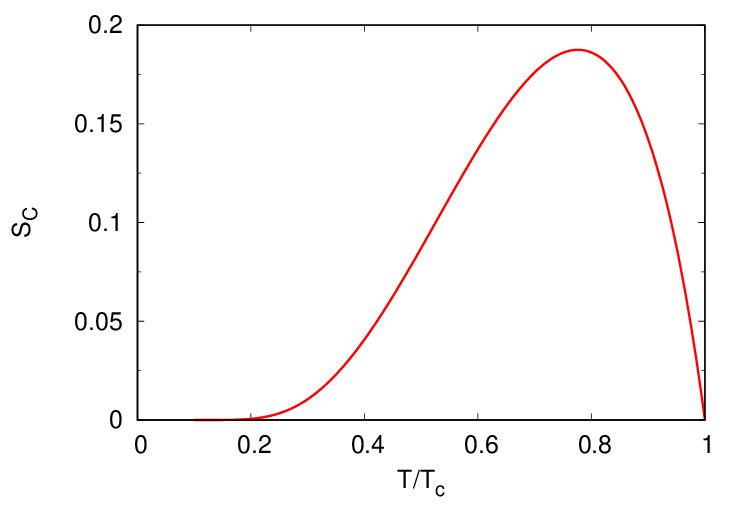}
\caption{Complexity of a system of spins using the Ising model at mean-field approximation in the absence of external magnetic field in the ferromagnetic phase, 
$\beta/\beta_c>1$, i.e. $T/T_c<1$.}
\label{fig.isingMF}
\end{figure}

Under such conditions the probability of magnetization $m$ is
\beq
p=\frac{1+m}{2},
\label{eq.pm}
\eeq
hence the complexity of the system per spin reads as a function already written in~Eq.(\ref{sstrB}). However, here $p$ is expressed in terms of the magnetization 
according to Eq.~(\ref{eq.pm}) where  $m$ is obtained as the solution of Eq.~(\ref{eq.msc}). The resulting curve as a function of $x=T/T_c$ 
is depicted in Fig.~\ref{fig.isingMF}. We have to emphasize that higher order fluctuations are suppressed within this approximation and also that it is applicable only
for high enough dimensions. So the complexity provides some further insight into the interplay of ferromagnetic ordering, $zJ$, and thermal fluctuations, $k_B T$.
It will be shown that approaching the critical temperature $T_c$, we find a special temperature, $T^*$ where the system shows maximal complexity.
As we can see the complexity apparently of the Ising model just close to the paramagnetic--ferromagnetic phase transition becomes maximal at a value of  $T^*\approx 0.776 T_c$.

\begin{figure}
\epsfxsize=8.4cm
\leavevmode
\epsffile{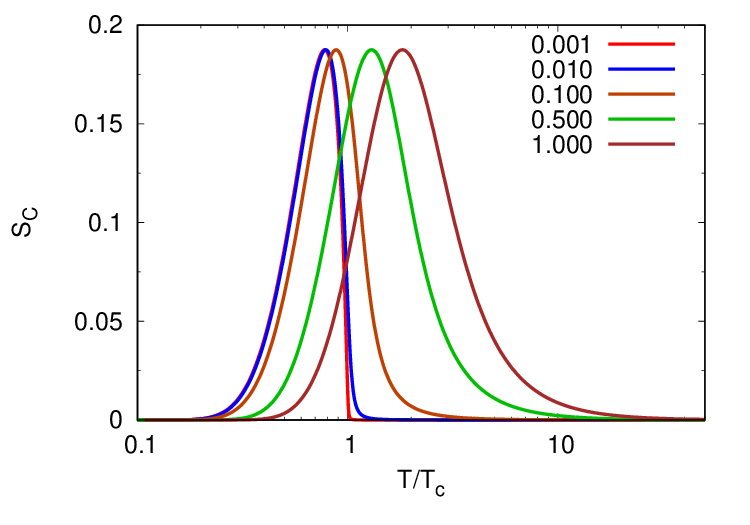}
\caption{Complexity of a system of spins using the Ising model at mean-field approximation for various values of the external magnetic field in the ferromagnetic phase, 
i.e. $x=\beta/\beta_c>1$ using a semilog plot. The curves are parametrized according to the ratio of ordering due to the magnetic field, $B$ and due to the coupling of
adjacent spins $zJ$ as $\alpha= \mu_B B/zJ$. The position of the maximum of the curves, $T^*/T_c$ increases with increasing $\alpha$.}
\label{fig.isingMF_}
\end{figure}

Finally let us investigate the same Ising problem for non-zero magnetic field, $B>0$. We may expect a stronger ordering tendency enforced by the presence of the
magnetic field. As long as $B=0$ there is a clear separation between the paramagnetic phase, when $\beta<\beta_c$, i.e $T>T_c$ with $m=0$ and the ferromagnetic phase, 
$\beta>\beta_c$, i.e. $T<T_c$ with $m=m_0\neq 0$. However, for nonzero external field, $B>0$ the magnetization is nonzero even for the paramagnetic phase. 
What we may compare is the complexity as a function of temperature parametrized by the ratio of magnetic coupling and ferromagnetic ordering, 
$\alpha=\beta_c\varepsilon_B=\varepsilon_B/k_B T_c=\varepsilon_B/zJ$. 
This is depicted in Fig.~\ref{fig.isingMF_}.
It is remarkable that Fig.~\ref{fig.isingMF_} shows parametrically very simple behavior for $T\ll T_c$, i.e. for low enough temperature the complexity is vanishing 
independent of the external field, whereas for large enough temperature the external field produces substantial effect and the complexity increases. However,
the temperature producing maximal complexity is also $B$ dependent and increases with increasing $B$. In this case there is no separate ferromagnetic and
paramagnetic phases and the system with maximal complexity is achieved at temperatures that increase beyond $T_c$ with increasing value of $\alpha$. The 
relation between $T^*/T_c$ and $\alpha$ is linear (not shown here) starting at $T^*/T_c\approx 0.776$ for $\alpha=0$, i.e. as $B=0$, but $T^*/T_c\propto\alpha$, 
hence for $\alpha\approx 0.2133$ we find $T^*\approx T_c$.

\section{Conclusions}

The understanding of two-level systems is nowadays more important than ever. They form the building blocks of future quantum computers and it is essential to
investigate how such systems behave under the influence of noise or disorder and a really appropriate and modern way to characterize the parameter dependence
is using some form of complexity measure. 

In the present work first we combined the notion of correlational entropy and statistical, entropic complexity in order to describe the behavior of two-level systems, or spins or
qubits. It has been shown that the effect of noise, i.e. disorder can be investigated in these systems and the complexity parameter used in the present work vanishes for
the trivial limiting cases of zero or large disorder while it becomes maximal when the interplay of the coupling to the noise (disorder) is maximal. 

In the second part we have shown that the complexity parameter introduced here can be applied for the characterization of the thermal equilibrium of random magnets in a general
paramagnetic phase or the Ising model at the mean-field approximation. We have shown that the complexity reveals interesting insights in these systems, as well. In both
cases, the interplay of thermal fluctuations driving towards maximal disorder and external magnetic field driving towards maximal order can be captured by the entropic
complexity. For the case of the Ising problem with external magnetic field it turns out that maximal complexity is achieved as a delicate interplay between thermal fluctuations, 
external magnetic field and the ferromagnetic coupling between adjacent spins.

In all the above cases maximal complexity of the quantum state is achieved whenever random, fluctuations and ordering are mutually important or otherwise the sate of the 
system is the most far away possible from the extremal cases of high ordering or high disorder.  How these characteristics depend on further degrees of freedom and how
special the state with maximal complexity is will be investigated in the future.

\begin{acknowledgements}
Project no. TKP2021-NVA-02 has been implemented with the support provided by the Ministry of Culture and Innovation of Hungary from the National Research, 
Development and Innovation Fund, financed under the TKP2021-NVA funding scheme. Additional support was provided by the Ministry of Culture and Innovation and the 
NRDI Office within the Quantum Information National Laboratory of Hungary (Grant No. 2022-2.1.1-NL-2022-00004).
\end{acknowledgements}

\end{document}